# An Alternative Cracking of The Genetic Code

Okunoye Babatunde O.

**Abstract**

We propose 22 'unique' solutions to the Genetic code: an alternative cracking, from the perspective of a Mathematician.

## INTRODUCTION

The Genetic code is now known to consist of triplet bases specifying for an amino acid (1). But before any biochemical appreciation of the genetic code, a page copy of DNA of any kind ordinarily looks like a cryptographic code – a discrete set of symbols produced by a stochastic process, which in this case are rows or lines of ten symbols: the purine and pyrimidine bases Adenine (A), Thymine (T), Guanine (G), and Cytosine (C).

In Communication theory a Secrecy system is defined as not one, but a set of many transformations of a language (2).

After the key is chosen only one of these transformations is used and one might be led from this to define a secrecy system as a single transformation on a language (2). In this paper, 22 'unique' solutions to the Genetic code are proposed, corresponding to enciphering with 22 different keys. DNA employed for this crypto-analytical attack is Bacteriophage T4 Genome (3) (168,900' – 137,280') in the 5' to 3' direction. The Genome was obtained from GenBank with Accession number AF158101.

**KEYS**

1. COMBINATION OF BASES AS KEY.

The DNA bases form number combinations with relative frequencies. The frequencies of English letters appearing in a page of a book (4) is recorded. Assuming our code is producing English text, we set up a simple substitution (2). Using the "Probable Word" method (2), we make out several words and phrases from the DNA cryptogram.

| BASE COMBINATIONN | ENGLISH LETTER | WORDS |
|---|---|---|
| 0055 | J | IT |
| 0028 | K | HE |
| 0118 | P | IS |
| 1117 | F | ME |
| 0037 | G | US |
| 0046 | Y | SANE |
| 0226 | B | DEN |
| 0127 | W | MEN |
| 0136 | M | ROB |
| 1144 | U | LIE |
| 0244 | L | SEED |
| 0145 | D | AREA |
| 1126 | R | HEAL |
| 0334 | O | WEST |
| 1333 | S | STAR |
| 2224 | N | ELITE |
| 1135 | T | BERTH |
| 0235 | H | HEARD |
| 1225 | I | TENET |
| 1135 | T | BERTH |
| 2233 | A | THEIR |
| 1234 | E | BEAR |

1. RANDOM PRIME NUMBER AS KEY.

DNA is now known to be a natural random number generator (5). Random numbers generated by DNA are crypto-analytical random key, except that enciphering with this key produces a secrecy system similar to enciphering with the first key – combination of bases.

Two secrecy systems R and S are defined to be "similar" if there exists a fixed transformation A with an inverse, A – 1, such that R = AS (2). If systems R and S are similar, a one-to-one correspondence between the cryptograms can be set up leading to the same *a posteriori possibilities,* cryptogram space, messages and transformations (2).

A new key, equivalent with a new set of transformations are the prime numbers within the random numbers. Prime numbers are the "atomic" elements of the natural numbers (6). The frequencies of the prime numbers are recorded, and a simple substitution is established with English letters based on frequencies/probabilities.

| PRIME NUMBER | ENGLISH LETTER | WORDS |
|---|---|---|
| 73 | M | TONE |
| 5 | U | SITE |
| 2 | L | SEES |
| 71 | D | HEED |
| 3 | R | SILT |
| 53 | O | TAUT |
| 11 | S | NEAT |
| 61 | N | TREE |
| 13 | T | DEIST |
| 43 | H | TAINT |

| 23 | I | AID |
| 31 | A | HAIR |
| 41 | E | NINE |

1. Tetrad Primes as Key.

   The distances between DNA segments with the combinations 55, 2233, 235, and 37 are recorded. A simple substitution is created with English letters.

| DISTANCE | ENGLISH LETTER | WORDS |
|---|---|---|
| 25 | Q | HE |
| 33 | X | OMIT |
| 36 | Z | TEA |
| 21 | J | EAST |
| 24 | V | AS |
| 20 | K | AT |
| 19 | P | DO |
| 38 | F | DEAN |
| 13 | G | AIM |
| 15 | Y | TO |
| 17 | C | EAT |
| 18 | B | KIN |
| 14 | W | TAN |
| 16 | M | ANT |
| 11 | U | AID |
| 12 | L | SEE |
| 10 | D | CASES |
| 9 | R | TALE |
| 8 | O | SINE |
| 7 | S | HORN |
| 6 | N | DINE |
| 5 | T | BETS |
| 4 | H | TEAR |
| 3 | I | BEE |
| 2 | A | NEW |
| 1 | E | ORE |

1. RECURSIVE STEPS / FUNCTION AS KEY.

   The frequencies of the distances between DNA segments with the recursive steps 0505, 2323, 1414 are recorded. A substitution table is created.

   | SOME WORDS |
   |---|
   | LENIN |
   | TEASE |
   | TEST |
   | SAID |
   | HOST |

2. CODE DEVELOPMENT BY STEPWISE REFINEMENT.

   Frequencies of the distances between the DNA segments with base combinations in the stepwise order 1 ? 2 ? 3 ? 4 are recorded and a substitution table created.

| SOME WORDS |
|---|
| HASTEN |
| DENT |
| TENET |
| EARN |
| AWE |

1. FIBONACCI NUMBERS AS KEY.

   Frequencies of the distances between DNA segments with base combinations consisting of Fibonacci numbers: 28, 55, 235 are recorded and a substitution table created.

| SOME WORDS |
|---|
| LET |
| HAT |
| HEN |
| RAN |

| MEN |
|---|
|  |

1. FIBONACCI SEQUENCE AS KEY.

   The frequencies of the distances between DNA segments, which can be arranged as Fibonacci sequences 235, 145, 055 are recorded, and a substitution table created.

   | SOME WORDS |
   |---|
   | MAR |
   | EEL |
   | RUM |
   | WOE |
   | DIRE |

2. LOOPS / LOOP STRUCTURES AS KEY.

   The frequencies of the distances between DNA segments with loops 2035, 2080, 1036, 0505 … are recorded and a substitution table created.

| SOME WORDS |
| --- |
| HEED |
| LOIN |
| HEAT |
| TENT |
| REAR |

1. LOOPS AS KEY (2).

    The frequencies of the distances between DNA segments which are repeated after a segment are recorded and a substitution table created. For example, **2035**, 1234, **3052** and **2323**, 0703, **3322** are considered loops.

| SOME WORDS |
| --- |
| WRAP |
| PILE |
| WET |
| GIN |

| DON |
| --- |

1. LOOPS AS KEY (3).

   The frequencies of the distances between zero's (0) in the cryptogram, i.e., base combinations are recorded and a substitution table created.

| SOME WORDS |
| --- |
| OWE |
| BET |
| SIN |
| ART |
| RAW |

2. STACKS / CLUSTERS AS KEY.

The frequencies of the distances between DNA segments which stack or cluster are recorded. 2323, **5050, 5500**, 1234

for example, is considered as stack or cluster. A substitution table is created.

| SOME WORDS |
|---|
| TRAIT |
| RIOT |
| ANNE |
| LOSS |
| NEAR |

1. QUEUES / LISTS AS KEY.

    The frequencies of the distances between DNA segments with the base combinations 1333, 2224, 0055, 1135, 2233, 1144, 3304, 4402 are recorded and a substitution table created.

| SOME WORDS |
|---|
| SHAM |
| AMOS |
| SANE |
| RAN |
| SEE |

1. CIPHER RATIO AS KEY.

   The ratio of the DNA bases A/T G/C and A + G / T+ C = 1 is exploited. The frequencies of the distances between 1 in the cryptogram (DNA segments) are recorded and a substitution table created.

| SOME WORDS |
|---|
| MASS |
| NUT |
| WIT |

| TROD |
|------|
| HOT  |

1. CIPHER RATIO AS KEY (2).

   The ratio of the number of cipher symbols (10) per segment of code and the type / class of cipher symbol (2) is used as key. The frequencies of the distances between 5 in the cryptogram are recorded and a substitution table created.

| SOME WORDS |
|------------|
| BIAS       |
| FRAY       |
| HADES      |
| DEN        |
| ODE        |

1. CIPHER CLASS / TYPE AS KEY.

    The cryptogram symbols are of two classes: Purine and Pyrimidine. The frequencies of the distances between 2 in the cryptogram are recorded and a substitution table created.

    | SOME WORDS |
    |---|
    | HEAT |
    | SET |
    | HERE |
    | TAB |
    | ORE |

2. CODON NUMBER AS KEY.

    The number of bases / symbols constituting a codon is employed as key. The frequencies of the respective distances between 3 in the cryptogram are recorded and a substitution table created.

| SOME WORDS |
|---|
| SEAT |
| YES |
| SAT |
| TEN |
| ASH |

1. CIPHER NUMBER AS KEY.

    The total number of different symbols (4), which constitute the cipher segment, is employed as key. The frequencies of the distances between 4 in the cryptogram are recorded and a substitution table created.

| SOME WORDS |
|---|
| SEAS |
| BAT |
| SEW |
| AGE |

| LIT |
|---|

1. CIPHER RATIO AS KEY (3).

The frequencies of the distances between DNA segments having the number 5 are recorded and a substitution table created. 5 represents the mid-point of the cipher. The related segments are 5005, 5023, 5113, 5014, 1225 and their various combinations.

| SOME WORDS |
|---|
| ASIA |
| HEATH |
| ALAN |
| MIRE |
| HEEL |

1. CIPHER DIVISIBILITY AS KEY.

   DNA segments A T G C form numbers. The frequencies of the distances between the numbers which can be divided by the number of symbols (4) are recorded and a substitution table created. As an example, for the base combination 1225, the numbers divisible by 4 are 1252, 2152, 2512 and 5212.

| SOME WORDS |
|---|
| FIESTA |
| NINTH |
| NET |
| TEASLE |
| BEN |

2. SORTING AS KEY (1).

The base combinations A T G C form numbers. The distances between a segment and the next segment with a higher numerical value is recorded. The frequencies of these distances are recorded and a substitution table created.

| SOME WORDS |
|---|
| MEW |
| RED |
| EON |
| ERA |
| ERE |

1. SORTING AS KEY (2).

   The base combinations A T G C are rotated and summed up. The frequencies of the resulting numbers are recorded and a simple substitution table created.

   A.

   1. 4510   11. 4213   21. 5212   31. 4420

   2. 2332   12. 2341   22. 3133   32. 2440

   3. 0721   13. 5113   23. 3421   33. 4312

4. 1522    14. 2512    24. 5212    34. 4222

5. 5311    15. 1612    25. 3421    35. 2413

6. 4312    16. 1342    26. 3610    36. 2404

7. 3412    17. 5311    27. 5320    37. 0433

8. 0442    18. 4213    28. 3115    38. 4501

9. 4213    19. 2512    29. 4510    39. 4033

10. 2215   20. 3412    30. 2251    40. 3502

41. 4222   51. 0523    61. 3502    71. 2332

42. 3223   52. 7201    62. 2512    72. 2332

43. 3304   53. 2512    63. 5122    73. 3223

44. 4213   54. 3412    64. 4213    74. 2413

45. 3520   55. 3313    65. 3313    75. 2530

46. 4312   56. 0451    66. 2431    76. 5023

B.

4201    4252   5335   4244

5372    2315   2142   4432

| | | | |
|---|---|---|---|
| 1322 | 1411 | 1321 | 2412 |
| 0212 | 3132 | 2312 | 0022 |
| 5430 | 1154 | 3353 | 2204 |
| 3344 | 6332 | 4631 | 4445 |
| 1114 | 1411 | 2121 | 1030 |
| 1222 | 2213 | 1005 | 3431 |
| 4243 | 2307 | 4232 | 4332 |
| 2222 | 5452 | 5255 | 0553 |
| 1122 | 1120 | 1501 | 3003 |
| 3523 | 2231 | 0122 | 3222 |
| 3434 | 6330 | 5432 | 3225 |
| 3253 | 1434 | 1234 | 2450 |
| 0121 | 2115 | 2113 | 2132 |
| 4302 | 1231 | 2331 | 3303 |

C.

| | | | |
|---|---|---|---|
| 4201 | 5372 | 1322 | 0212 |
| 4252 | 2315 | 1411 | 3132 |
| 5335 | 2142 | 1321 | 2312 |
| 4244 | 4432 | 2412 | 0022 |
| | | | |
| 5430 | 3344 | 1114 | 1222 |
| 1154 | 6332 | 1411 | 2213 |
| 3353 | 4631 | 2121 | 1005 |
| 2204 | 4445 | 1030 | 3431 |
| | | | |
| 4243 | 2222 | 1122 | 3523 |
| 2307 | 5452 | 1120 | 2231 |
| 4232 | 5255 | 1501 | 0122 |
| 4332 | 0553 | 3003 | 3222 |
| | | | |
| 3434 | 3253 | 0121 | 4302 |
| 6330 | 1434 | 2115 | 1231 |
| 5432 | 1234 | 2113 | 2331 |
| 3225 | 2450 | 2132 | 3303 |

D.

| | | | |
|---|---|---|---|
| 4201 | 5430 | 4243 | 3434 |
| 4252 | 1154 | 2307 | 6330 |
| 5335 | 3353 | 4232 | 5432 |
| 4244 | 2204 | 4332 | 3225 |
| 5372 | 3344 | 2222 | 3253 |
| 2315 | 6332 | 5452 | 1434 |
| 2142 | 4631 | 5255 | 1234 |
| 4432 | 4445 | 0553 | 2450 |
| 1322 | 1114 | 1122 | 0121 |
| 1411 | 1411 | 1120 | 2115 |
| 1321 | 2121 | 1501 | 2113 |
| 2412 | 1030 | 3003 | 2132 |
| 0212 | 1222 | 3523 | 4302 |
| 3132 | 2213 | 2231 | 1231 |
| 2312 | 1005 | 0122 | 2331 |
| 2412 | 1030 | 3003 | 2132 |

| 0212 | 1222 | 3523 | 4302 |
| --- | --- | --- | --- |
| 3132 | 2213 | 2231 | 1231 |
| 2312 | 1005 | 0122 | 2331 |
| 0022 | 3431 | 3222 | 3303 |

E.

| 7 | 12 | 13 | 14 |
| --- | --- | --- | --- |
| 13 | 11 | 12 | 12 |
| 16 | 14 | 11 | 14 |
| 14 | 8 | 12 | 12 |

| 17 | 17 | 8 | 13 |
| --- | --- | --- | --- |
| 11 | 14 | 16 | 12 |
| 9 | 14 | 17 | 10 |
| 13 | 17 | 13 | 11 |

| 8 | 7 | 6 | 4 |
| --- | --- | --- | --- |
| 7 | 7 | 4 | 9 |
| 7 | 6 | 7 | 7 |
| 9 | 4 | 6 | 8 |
| 5 | 7 | 13 | 9 |
| 9 | 8 | 8 | 7 |
| 8 | 6 | 5 | 7 |
| 4 | 11 | 9 | 9 |

| SOME WORDS |
|---|
| INN |
| SHUN |
| LET |
| RILE |
| ROE |

1. MATRIX DETERMINANT AS KEY.

   DNA segments are viewed as 2 × 2 matrix of the form: A C

   G T

Where A T G C respectively are the numbers of the bases per segment and the determinants resolved. Only the absolute values are used. The frequency of the numbers is recorded and a simple substitution table is created.

| SOME WORDS |
| --- |
| TIES |
| DO |
| END |
| TAT |
| SO |

**CONCLUSION AND DISCUSSION**

Using the "Probable Word" method (2), we discover a plethora of words and phrases underneath the cryptogram, sometimes just enough to form elementary and loose sentences. The assumption made in this work is that the DNA cryptogram is producing English Text. It remains to be seen however if similar results will be obtained using other languages.

This work presents an interesting puzzle, because it suggests that the genetic code, in addition to encoding biochemical information, also encodes information 'readable' to the human mind – linguistic information. On the other hand, it could be dismissed outrightly – like the case of the man who returns home to find his front door ajar and exclaims: My God! The Martians have been here! It's not wrong, just highly speculative.